\documentclass{ws-ijmpd}
\usepackage{epsfig}
\usepackage{epstopdf}
\usepackage{color}

\begin{document}
	
\title{Compact stellar models in modified gravity}

\author{Ines G. Salako}
\address{Ecole de G\'enie Rural (EGR), 01 BP 55 K\'etou, Benin\\Institut de Math\'ematiques et de Sciences Physiques  (IMSP), 01 BP 613  Porto-Novo, Benin\\Ecole Doctorale des Sciences Exactes et Appliqu\'ees, UAC, B\'enin\\ inessalako@gmail.com}

\author{Chayan Ranjit}
\address{Department of Mathematics, Egra S.S.B. College, Purba Medinipur 721429, West Bengal, India\\chayanranjit@gmail.com}

\author{M. Khlopov}
\address{National Research Nuclear University, MEPHI (Moscow Engineering Physics Institute), Moscow 115409, Russia\\Universit\'e de Paris, CNRS, Astroparticule et Cosmologie, F-75013 Paris, France\\Institute of Physics, Southern Federal University, Rostov on Don 344090, Russia\\khlopov@apc.in2p3.fr}

\author{Saibal Ray\footnote{Corresponding Author}}
\address{Department of Physics, Government College of Engineering and Ceramic Technology, Kolkata 700010, West Bengal, India\\ saibal@associates.iucaa.in}

\author{Utpal Mukhopadhyay}
\address{Satyabharati Vidyapith, Nabapally, Barasat, North 24 Parganas, Kolkata 700126, West Bengal, India\\utpalsbv@gmail.com}
	
\maketitle	

\abstract{In the present investigation compact stellar models are dealt with in the framework of the modified gravity theory, specifically of $f(\mathbb{T},\mathcal{T})$ type. We have considered that the compact objects are following a spherically symmetric static metric and obtained the Einstein field equations in the spacetime of $f(\mathbb{T},\mathcal{T})$. To make the Einstein equations solvable we employ the methodology of conformal Killing vectors. Thereafter by using the MIT bag equation of state to the compact stars, considering that the stars are formed by strange quark, we find the solutions set. The solutions are examined via several physical tastings which exhibit viability of the model.  }


\section{Introduction}
More than a century has elapsed since Schwarzschild~\cite{Schwarzschild1916a,Schwarzschild1916b} found the first ever solution of the Einstein Field Equations (EFE) for a spherically symmetric, static, isotropic, uncharged fluid distribution immediately after the publication of Einstein's General Theory of Relativity (GTR). Afterwards, during this long period, various other solutions of EFE including the following two famous~\cite{Oppenheimer1939,Tolman1939} were obtained and it was found~\cite{Delgaty1998} that for static, spherically symmetric configuration, only 16 out of 127 solutions are acceptable. It has been also proved beyond doubt that GTR is a formidable weapon for tackling various astrophysical as well as cosmological issues where gravity plays a dominant role. In the late stage evolution of stars, GTR is used since due to abnormally high density of stellar body at this stage, gravity becomes the main force to reckon with. Previously it was thought that white dwarf, neutron star and black hole were the only three possible options through which a star could end its stellar life. But, now it has been found that strange quark star can also be a stellar corpse in the late stage evolution through conversion of neutrons to $u$, $d$ and $s$ quarks ~\cite{Bombaci2004,Staff2007,Herzog2011}, of which mostly are strange quarks. Apart from quarks, some electron type leptons may be present in stars made of quark matter. It may be mentioned that quark matter is more stable than ordinary matter and very massive neutron stars usually reduce to black holes whereas low mass neutron stars may become quark stars. 

Although GTR has been very successful in explaining a number of physical phenomena and till now it is used by majority of researchers for their investigations in cosmological and astrophysical problems, however this theory is not totally free from its limitations. Apart from its inherent problem of singularity~\cite{Wheeler1962}, GTR cannot explain properly the present observational evidence of accelerating universe without invoking the idea of dark energy and dark matter whose exact entity is unknown till date~\cite{Kamenschik2001,Padmanabhan2002,Bento2002,Caldwell2002,Nojiri2003a,Nojiri2003b,Riess2004,Eisenstein2005,Astier2006,Spergel2007}. To alleviate this problem, in recent years, some modifications have been made in the geometrical part of EFE. This means that the Einstein-Hilbert action is modified by introducing a generalised functional form of the Lagrangian density. In recent years, this new gravitational theory, named as modified gravity, has been used for addressing various issues in the gravitational researches. There are many variants of modified gravity, viz. $f(R)$ gravity~\cite{Nojiri2003,Carroll2004,Allemandi2003,Nojiri2007,Bertolami2007}, $f(\mathcal{T})$ gravity~\cite{Bengachea2009,Linder2010}, $f(G)$ gravity~\cite{Bamba2010a,Bamba2010b,Rodrigues2010}, $f(R,\mathcal{T})$ gravity~\cite{Harko2011,Harko2008,Bisabr2012,Jamil2012,Alvarenga2013,Shabani2013,Shabani2014,Zaregonbadi2016,Shabani2017a,Shabani2017b}, $f(\mathbb{T},\mathcal{T})$ gravity~\cite{Junior2014,Salako2015,Ganiou2016a,Ganiou2016b,Salako2017,Ghosh2020}, $f(R,G)$ gravity~\cite{Nojiri2005} etc. where $R$, $\mathbb{T}$, $\mathcal{T}$ and $G$ are scalar curvature, torsion scalar, trace of the energy-momentum tensor and Gauss-Bonnet scalar respectively.  
          
Starting from the work of Schwarzschild~\cite{Schwarzschild1916a,Schwarzschild1916b} previously only isotropic stellar models were considered. But, Ruderman~\cite{Ruderman1972} first demonstrated that anisotropy can develop in a star with density $10^{15}$~gm/cc or more. After this seminal work Herrera et al.~\cite{Herrera1997} studied the nature of anisotropy in a self-gravitating system. Anisotropy means that the radial and tangential pressures in a star are different which can affect the energy density (and hence space-time geometry), the total mass, the gravitational redshift and the frequency of the fundamental mode of compact stars. Anisotropic stellar models have been investigated by several workers within the framework of GTR~\cite{Hossaein2012,Kalam2014,Rahaman2014,Bhar2015,Abbas2015,Abranil2016,Murad2016} as well as by using modified gravity~\cite{Oliveira2015,Sarif2018,Saha2018,Maurya2019,Prasad2019,Saha2019,Abbas2019,Shahzad2019,Naxar2020}. Rahaman et al.~\cite{Rahaman2012} have studied anisotropic strange star using GTR and Deb et al.~\cite{Deb2018a,Deb2018b} have studied anisotropic as well as isotropic strange stars in $f (R,\mathcal{T})$ gravity. Recently, Salako et al.~\cite{Salako2020}, have studied anisotropic strange star model using $f(T,\mathcal{T})$ gravity. So, the present investigation can be regarded as a complementary work of the paper of Salako et al.~\cite{Salako2020}.  Both the works have been performed for strange star model under $f(\mathbb{T},\mathcal{T})$ gravity but in the present paper we are considering isotropic fluid sphere which is treated with the method of conformal Killing vectors to solve the EFE in a more efficient way.
          
Choice of equation of state (EOS) is an important factor for cosmological as well as astrophysical investigations. MIT bag model is a favourite choice of researchers as an EOS. It has been used by several workers within GTR~\cite{Abranil2016,Murad2016,Rahaman2014} and modified gravity~\cite{Rahaman2012,Deb2016,Shahzad2019}. The bag constant involved in this EOS affects the energy-momentum tensor and hence the space-time geometry of the star as well~\cite{Chodos1974}. For this reason, in spite of its {\it ad hoc} nature, this EOS has been used extensively. In the present work, EOS of MIT bag model is used along with the usual pressure density relation $p = \omega \rho$ for compact stars.
            
The present work is organized as follows. In Section 2, basic mathematical formalism of $f (\mathbb{T},\mathcal{T})$ gravity and formulation of the field equations have been done. In Section 3, the field equations are solved to determine parameters, like $p_{eff}$, $\rho_{eff}$, $\nu(r)$, $\lambda(r)$ and $m(r)$, while various physical features, viz. fulfillment of the energy condition, mass-radius relation, stability analysis etc. are presented in several subsections of Section 4. The Section 5 is exclusively for some concluding remarks.

\section{Einstein's field equations in $f(\mathbb{T}, \mathcal{T})$ gravity} \label{sec4}
The Einstein field equations for $f(\mathbb{T},\mathcal{T})$ reduces to~\cite{Harko2014}
\begin{equation}
G_{\mu\nu}=8\pi T_{\mu\nu}^{eff},\label{eq201}
\end{equation}
where
\begin{eqnarray}
T_{\mu\nu}^{eff}= g_{\mu\nu} \Bigg[\frac{\Big(-\varpi\,(\rho -3 p) +2  \Lambda \Big)}{16\pi }
+\frac{\varpi p}{8\pi }  \Bigg]+ T_{\nu\mu} \Bigg(1+ \frac{\varpi}{8\pi } \Bigg).
\end{eqnarray}

Also, the covariant derivative of Eq. (\ref{eq201}) can be provided as
\begin{eqnarray}
\nabla_{\mu} T_{\nu}^{\;\;\mu}= \frac{1}{\Big(4\pi +(1/2)\varpi\Big)}\Bigg\{\frac{\varpi}{4}\left(\partial_{\nu}\mathcal{T}\right)-\frac{\varpi}{2} \partial_{\nu}p_{}\Bigg\}. \label{eqnonconserv}
\end{eqnarray}

Hence under the line element of spherically symmetric and static spacetimes 
\begin{eqnarray}\label{3.13}
 ds^2=e^{\nu(r)}dt^2-e^{\lambda(r)}dr^2-r^2(d\theta^2+\sin^2\theta d\phi^2)
 \end{eqnarray}
the modified Einstein field equations and the torsion scalar can be written as~\cite{Salako2020}
\begin{equation}
G_0^{0}=\frac{e^{-\lambda}}{r^{2}}(-1+e^{\lambda}+\lambda'r),\label{eq28}
\end{equation}

\begin{equation}
G_1^{1}=\frac{e^{-\lambda}}{r^{2}}(-1+e^{\lambda}-\nu'r),\label{eq29}
\end{equation}

\begin{equation}
G_2^{2}=G_3^{3}=\frac{e^{-\lambda}}{4r}[2(\lambda'-\nu')-(2\nu''+\nu'^{2}-\nu'\lambda')r],\label{eq30}
\end{equation}

\begin{eqnarray}
T(r) &=& \frac{2e^{-\lambda}}{r^2}\left(e^{\lambda/2}-1\right)\left(e^{\lambda/2}-1-r \nu^{\prime}\right)\label{te},
\end{eqnarray}
where the prime  ($^{\prime}$) denotes the derivative with respect to  the radial coordinate $r$.

From the above relationships we find the explicit form of the Einstein field equation \eqref{eq201} as follows:
\begin{eqnarray}\label{3.21}
& \hspace{-1cm} {{\rm e}^{-\lambda }} \left( {\frac {\lambda^{{\prime}}}{r}}-\frac{1}{r^2}\right) +\frac{1}{r^2}= 8\pi \Bigg\{  \Bigg[\frac{\Big(-\varpi\,(\rho -3 p) +2  \Lambda \Big)}{16\pi }
+\frac{\varpi p}{8\pi }  \Bigg]+ \rho\Bigg(1+ \frac{\varpi}{8\pi } \Bigg)  \Bigg\} =8\pi \rho^{\textit{eff}}, \\ \label{3.22}
& \hspace{-1cm} {{\rm e}^{-\lambda}} \left( {\frac {\nu^{{\prime}}}{r}}+\frac{1}{r^2}\right) -\frac{1}{r^2}=-8\pi \Bigg\{\Bigg[\frac{\Big(-\varpi\,(\rho -3 p) +2  \Lambda \Big)}{16\pi }
+\frac{\varpi p}{8\pi }  \Bigg]-p \Bigg(1+ \frac{\varpi}{8\pi } \Bigg)  \Bigg\} =8\pi  p^{\textit{eff}},
\end{eqnarray}
where the primes denote  the differentiation with respect to the radial coordinate $r$. Here $\rho^{\textit{eff}}$ and $p^{\textit{eff}}$, the effective density and pressure of the matter distribution, respectively, are given by
\begin{eqnarray}\label{3.23}
& \rho^{\textit{eff}}=\rho + \frac{\varpi \rho}{16\pi } + \frac{5\varpi\,p}{16\pi } + \frac{\Lambda }{4\pi },
\\ \label{3.23a}
& p^{\textit{eff}}=p + \frac{\varpi \rho}{16\pi } - \frac{3\varpi\,p}{16\pi } - \frac{\Lambda }{4\pi }.
\end{eqnarray}

We assume that the SQM distribution inside the strange stars is governed by the simple phenomenological MIT Bag model EOS~\cite{Chodos1974}. In bag model, by introducing {\it ad hoc} bag function all the corrections of energy and pressure functions of SQM have been maintained. We also consider that the quarks are non-interacting and massless in a simplified bag model. The quark pressure therefore can be defined as
\begin{equation}\label{2.8}
{p}={\sum_{f=u,d,s}}{p^f}-{B},
 \end{equation}
where $p^f$ is the individual pressure of the up~$\left(u\right)$, down~$\left(d\right)$ and strange~$\left(s\right)$ quark flavors and $B$ is the vacuum energy density (also well known as Bag constant) which is a constant quantity within a numerical range. In the present article we consider the value of Bag constant as $B=83~MeV/{{fm}^3}$~\cite{Rahaman2014,Aziz2019,Jasim2020,Biswas2021}.

Now the individual quark pressure ($p^f$) can be defined as $p^f=\frac{1}{3}{{\rho}^f}$, where ${{\rho}^f}$ is the energy density of the individual quark flavor. Hence, the energy density, $\rho$ of the de-confined quarks inside the bag is given by
\begin{equation}
{{\rho}}={\sum_{f=u,d,s}}{{\rho}^f}+B. \label{2.9}
\end{equation}

Using Eqs.~(\ref{2.8})~and~(\ref{2.9}) we have the EOS for SQM given as
\begin{equation}
{p}=\frac{1}{3}({{\rho}}-4B).\label{2.10}
\end{equation}

It is observed that ignoring critical aspects of the quantum particle physics in the framework of GR several authors~\cite{Zdunik2000,Maieron2004,Nicotra2006,Bao2009,Uechi2010,Isayev2015,Cardoso2017,Joshi2020} successfully have been introduced this simplified form of the MIT Bag EOS to study stellar systems made of SQM.

To have non-singular monotonically decreasing matter density inside the spherically symmetric stellar system, following Mak and Harko~\cite{Harko2002}, we assume simplified form of $\rho$ given as
\begin{equation}\label{2.11}
\rho(r)=\rho_c\left[1-\left(1-\frac{\rho_0}{\rho_c}\right)\frac{r^{2}}{R^{2}}\right],
\end{equation}
where $\rho_c$ and $\rho_0$ are constants and denote the maximum and minimum values of $\rho$ at the center and on the surface, respectively.

We define the mass function of the spherically symmetric stellar system as
\begin{equation}\label{2.12}
m \left( r \right) =4\,\pi\int_{0}^{r}\!{{\rho}_{eff}} \left( r \right) {r}^{2}{dr}.
\end{equation}

At this juncture we consider the Schwarzschild metric to represent the exterior spacetime of our system given as
\begin{eqnarray}\label{2.13}
 {ds}^2=\left(1-\frac{2M}{r}\right)dt^2- \frac{{dr}^2}{\left(1-\frac{2M}{r}\right)}-r^2(d\theta^2+\sin^2\theta
d\phi^2),\nonumber\\
 \end{eqnarray}
where $M$ is the total mass of the stellar system.

Now, substituting Eq.~(\ref{2.12}) into Eq.~(\ref{3.21}) we find
\begin{eqnarray}\label{2.14}
 {{\rm e}^{-\lambda \left( r \right) }}=1-{\frac {2m}{r}}.
 \end{eqnarray}

Also the conservation equation (\ref{eqnonconserv})in $f(\mathbb{T},\mathcal{T})$ gravity takes the form as follows
\begin{equation}
 p'+\frac{\nu'}{2} (p+\rho)=\frac{1}{\Big(4\pi +(1/2)\varpi\Big)}\Bigg\{\frac{\varpi \rho'}{4}-\frac{5\varpi p'}{4} \Bigg\}.\label{eq40}
\end{equation}

The essential stellar structure equations required to describe static and charged spherically symmetric sphere
in $f(R, \mathcal{T})$ gravity theory are given by
\begin{eqnarray}\label{3.24}
 \frac{dm}{dr}&=&4\pi r^2\, \rho^{\textit{eff}},\,\cr \label{3.25}
\frac{dp}{dr}&=&\frac{1}{\left[1+\frac{\varpi}{16\pi+2\varpi}\left(5-\frac{d\rho}{dp}\right)\right]}\Bigg\lbrace -\left(\rho+p\right)\bigg[\frac{\Big\lbrace 4\pi\, r\,p^{\textit{eff}}+\frac{m}{r^2}  \Big\rbrace }{\left(1-\frac{2m}{r}\right)}
\bigg] \Bigg\rbrace.
\end{eqnarray}

\section{Solution to the Einstein field equation in $f(\mathbb{T}, \mathcal{T})$ gravity}\label{sec5}
The conformal Killing vector (CKV) is defined as
\begin{equation}\label{4.1A}
\textbf{L}_{\xi}g_{ij}=\xi_{i;j}+\xi_{j;i}=\psi g_{ij},
\end{equation}
where $\bf{L}$ is the Lie derivative operator, which describes the interior gravitational field of a star with respect to the vector
field $\xi$ and $\psi$ is the conformal factor.

Let us consider that our static spherically symmetric spacetime admits an one-parameter group of conformal motion in above
mentioned framework as background and also consider the metric given by Eq. (\ref{3.13}) which is conformally mapped onto
itself along $\xi$.

Now from Eqs. (\ref{4.1A}), we have
\begin{equation}\label{4.2A}
\textbf{L}_{\xi}g_{ik}=\xi_{i;k}+\xi_{k;i}=\psi g_{ik},
\end{equation}
where $\xi_i=g_{ik}\xi^{k}$, from which we find the following expression as follows:
\begin{eqnarray}\label{4.3A}
 \xi^{1}\nu^{\prime}=\psi,\cr
\xi^{1}\lambda^{\prime}+2\xi^{1}_{;1}=\psi ,\cr \xi^{1}=\frac{\psi
r}{2},\cr \xi^{2}=\text{constant},
\end{eqnarray}
where 1 and 2 stand for $r$ and $\theta$ respectively.

From the above set of equations, we get
\begin{equation}\label{4.4A}
e^{\nu}=C_2^2 r^2,
\end{equation}

\begin{equation}\label{4.5A}
e^{\lambda}=\frac{C_3^2}{\psi^2},
\end{equation}

\begin{equation}\label{4.6A}
\xi^{i} =C_1 \delta_{2}^i+\left[\frac{\psi
r}{2}\right]\delta_1^{i},
\end{equation}
where $C_1$, $C_2$ and $C_3$ all are integrating constants.

Now, from Eqs.(\ref{3.21}) using Eqs.(\ref{4.5A}) we have
\begin{equation}\label{4.7A}
-\frac{2\psi\psi^{\prime}}{r C_3^2}-\frac{\psi^2}{r^2
C_3^2}+\frac{1}{r^{2}}=8\pi \rho^{\textit{eff}},
\end{equation}
where $\rho^{\textit{eff}}=\rho + \frac{\varpi \rho}{16\pi } + \frac{5\varpi\,p}{16\pi } + \frac{\Lambda }{4\pi }$.

On the other hand, from Eqs.(\ref{3.22}) using Eqs. (\ref{4.4A}) and (\ref{4.5A}) we have
\begin{equation}\label{4.8A}
\frac{3\psi^{2}}{r^2 C_3^2}-\frac{1}{r^2}=8\pi p^{\textit{eff}},
\end{equation}
where $p^{\textit{eff}}=p + \frac{\varpi \rho}{16\pi } - \frac{3\varpi\,p}{16\pi } - \frac{\Lambda }{4\pi }.$

To solve Eqs. (\ref{4.7A}) and (\ref{4.8A}) let us assume the equation state of fluid of normal matter as
\begin{equation}\label{4.9A}
p=\omega\rho,
\end{equation}
where $\omega$ ($0<\omega<1$) is the equation of the state parameter. 

Therefore, from Eqs. (\ref{4.7A}) and (\ref{4.8A}) with the help of Eq. (\ref{4.9A}) we get
\begin{equation}\label{4.10A}
\rho=\frac{1}{8\pi\epsilon_1}\left[-\frac{2\psi\psi^{\prime}}{r
C_3^2}-\frac{\psi^2}{r^2 C_3^2}+\frac{1}{r^{2}}-2\Lambda\right]
\end{equation}
and
\begin{equation}\label{4.11A}
\rho=\frac{1}{8\pi\epsilon_2}\left[\frac{3\psi^{2}}{r^2
C_3^2}-\frac{1}{r^2}+2\Lambda\right],
\end{equation}
where
$\epsilon_1=1+\frac{\varpi}{16\pi}+\frac{5\omega\varpi}{16\pi} \neq 0$ and $\epsilon_2=\omega+\frac{\varpi}{16\pi}-\frac{3\omega\varpi}{16\pi} \neq 0$.

Now, equating the above two expressions of the density $\rho$ we have found the following differential equation in $\psi$ as follows:
\begin{equation}\label{4.12A}
2r\psi\psi^{\prime}+\left(1+3\frac{\epsilon_1}{\epsilon_2}\right)\psi^2
=\left(1+\frac{\epsilon_1}{\epsilon_2}\right)C_3^2-
2\left(1+\frac{\epsilon_1}{\epsilon_2}\right)\Lambda C_3^2 r^2.
\end{equation}

Solving one can get the following solution of $\psi^2$ as:
\begin{equation}\label{4.13A}
\psi^{2}=C_3^2(\epsilon_1+\epsilon_2)\left[\frac{1}{3\epsilon_1+\epsilon_2} - \frac{2\Lambda r^2}{3\epsilon_1+3\epsilon_2}\right],
\end{equation}
where
$\epsilon_1=1+\frac{\varpi}{16\pi}+\frac{5\omega\varpi}{16\pi}$ and $\epsilon_2=\omega+\frac{\varpi}{16\pi}-\frac{3\omega\varpi}{16\pi}$.

Now from Eqs. (\ref{4.4A}) and (\ref{4.5A}) using Eq. (\ref{4.13A}), we get
\begin{equation}\label{4.14A}
\nu(r)=2 \textit{ln}(C_2 r)
\end{equation}
and
\begin{equation}\label{4.15A}
\lambda(r)=2 \textit{ln}[C_3]-\textit{ln}\left[-\frac{C_3^2
\left(8 \pi  \left(-3 (1+\omega )+2 r^2 \Lambda  (3+\omega
)\right)+\left(-3 (1+\omega )+4 r^2 (\Lambda +3 \Lambda
 \omega )\right) \varpi\right)}{6 \left(4 \pi  (3+\omega )
 +(1+3 \omega ) \varpi\right)}\right].
\end{equation}

\begin{figure}[!htpb]\centering
\includegraphics[width=6cm]{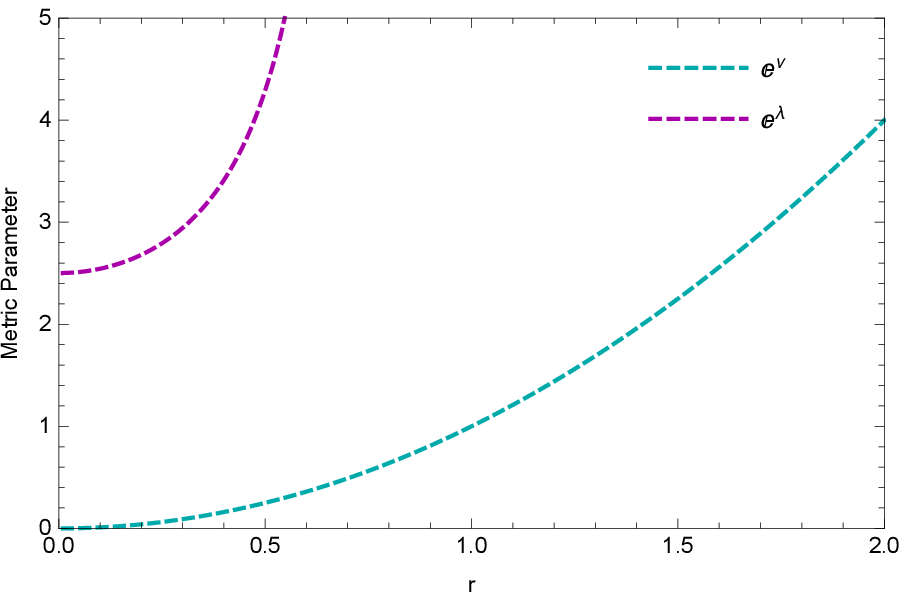}
\includegraphics[width=6cm]{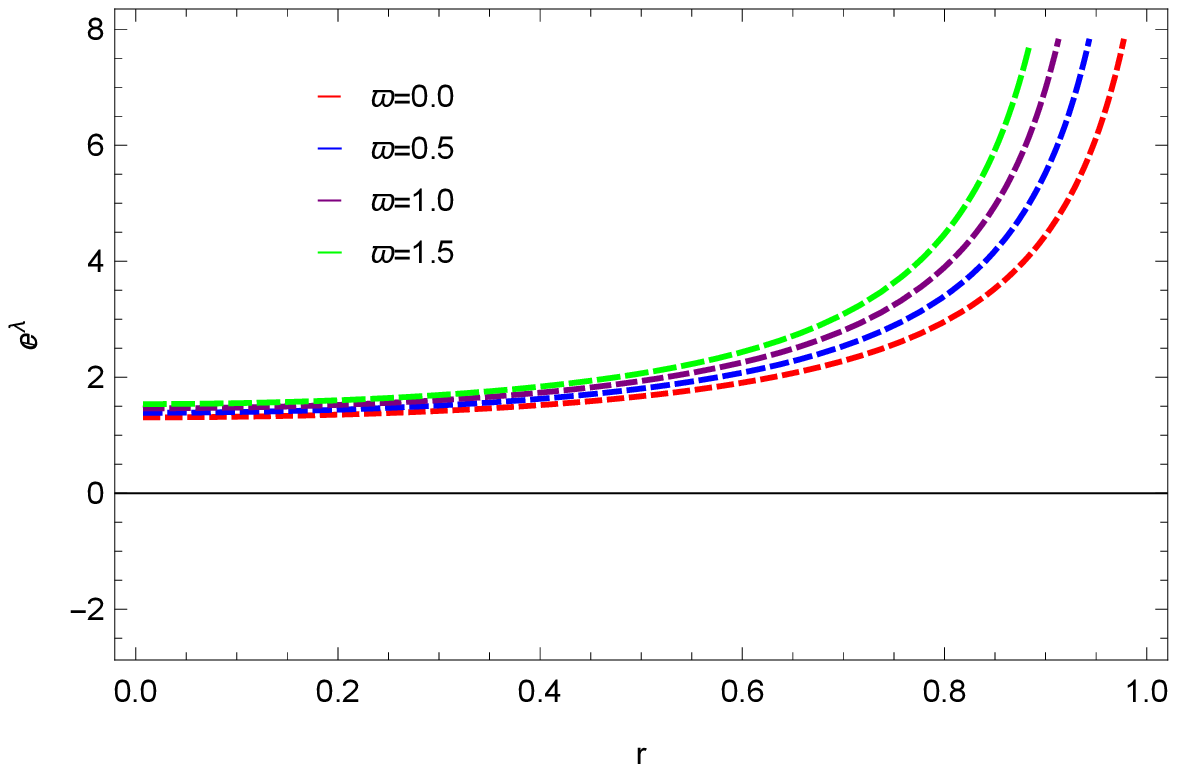}
\caption{Variation of the metric potentials $e^{\nu}$ and $e^{\lambda}$ with respect to $r$ (left panel) while variation of the metric potential $e^{\lambda}$ for $\varpi=0.0,0.5,1.0,1.5$ (right panel)} \label{fig1}
\end{figure}

Again substituting the value of Eq. (\ref{4.13A}) in Eqs. (\ref{4.7A}) and (\ref{4.8A}), we have
\begin{equation}\label{4.14A}
\rho^{\textit{eff}}=\frac{16 \pi  \left(1+r^2 \Lambda  (3+\omega
)\right)+\left(1+5 \omega +4 r^2 (\Lambda +3 \Lambda  \omega
)\right) \varpi}{16 \pi  r^2 \left(4 \pi  (3+\omega )+(1+3 \omega
) \varpi\right)}
\end{equation}
\begin{equation}\label{4.14A}
p^{\textit{eff}}=-\frac{16 \pi  \left(-\omega +r^2 \Lambda
(3+\omega )\right)+\left(-1+3 \omega +4 r^2 (\Lambda +3 \Lambda
\omega )\right) \varpi}{16 \pi  r^2 \left(4 \pi (3+\omega )+(1+3
\omega ) \varpi\right)}.
\end{equation}

\begin{figure}[!htpb]\centering
\includegraphics[width=6cm]{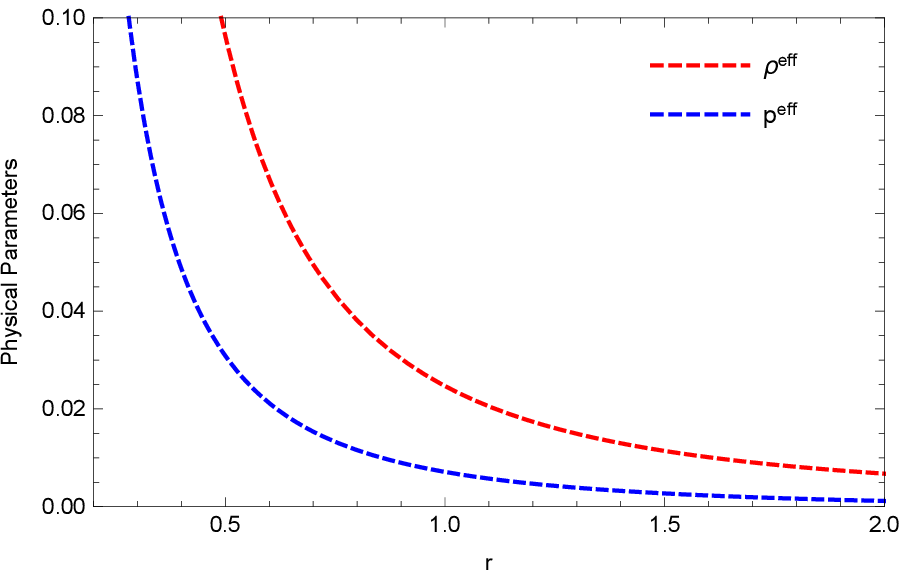}\\
\includegraphics[width=6cm]{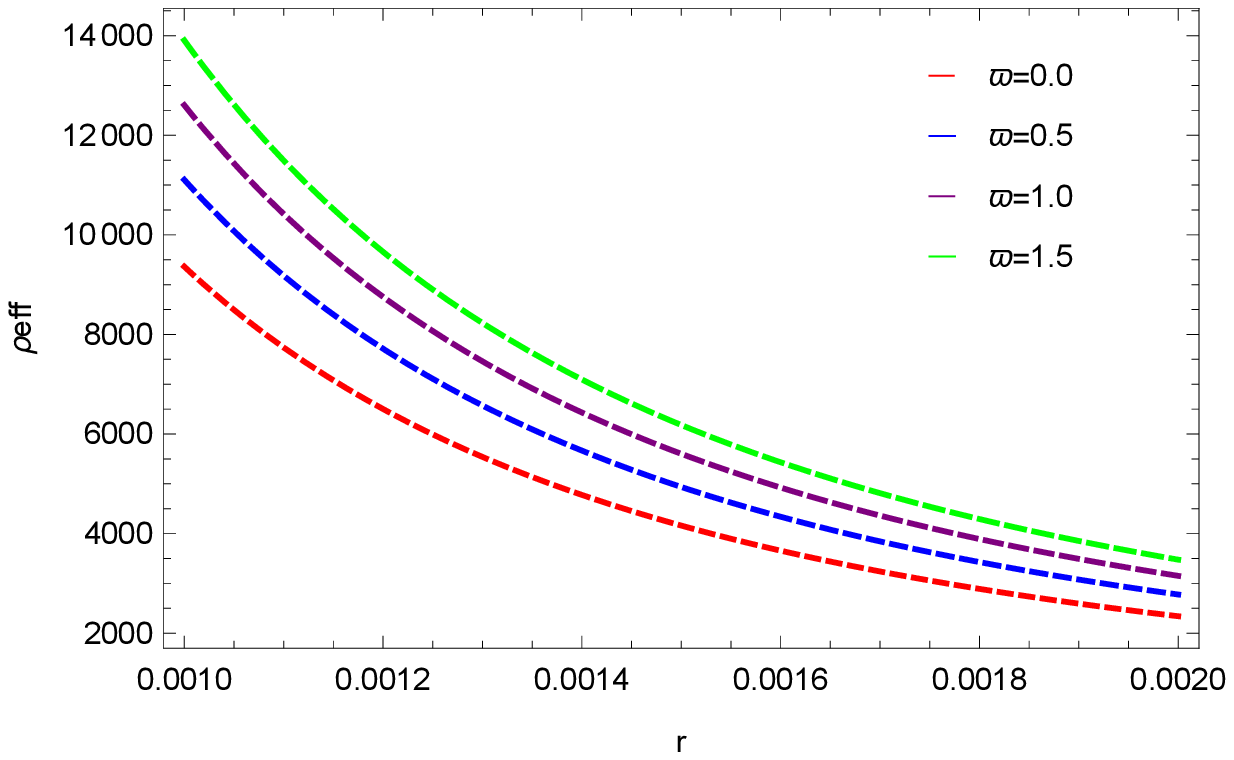}
\includegraphics[width=6cm]{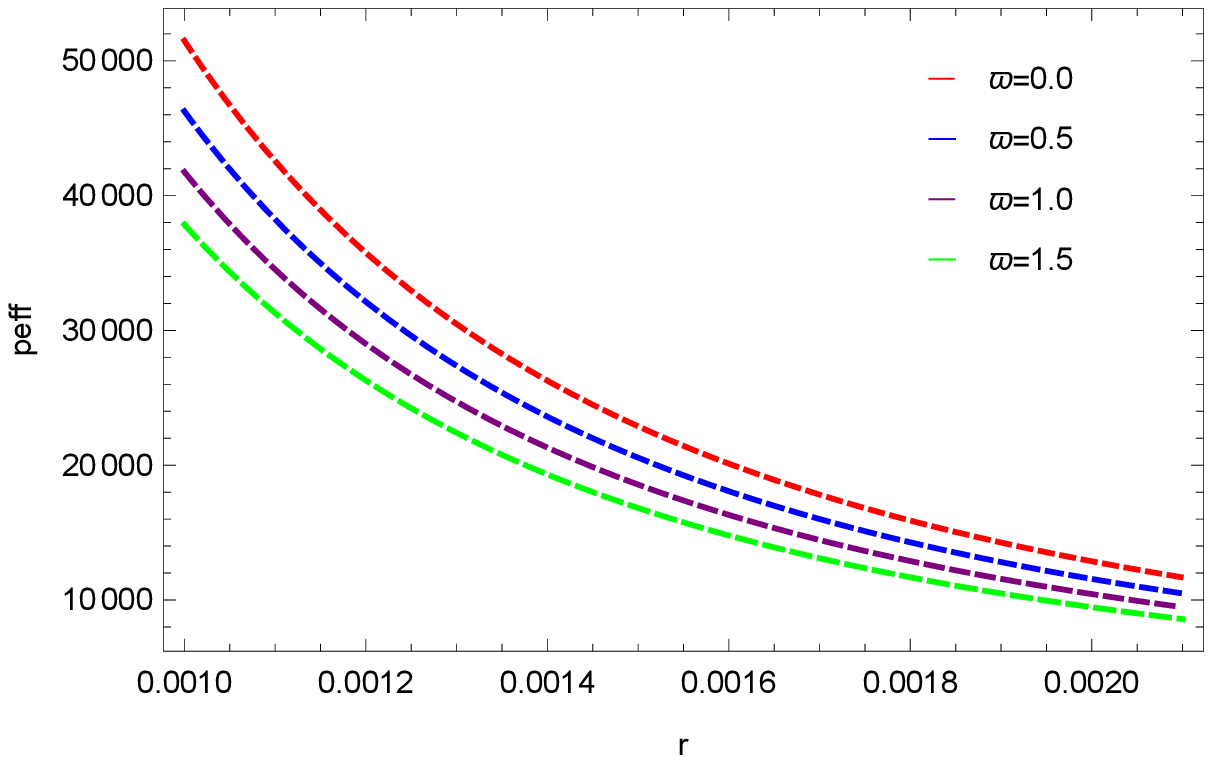}
\caption{Variations of the density $\rho^{eff}$ ($km^{-2}$) and pressure $p^{eff}$ ($km^{-2}$) with respect to the
radial coordinate $r$ (km) (upper panel) while their variations for $\varpi=0.0,0.5,1.0,1.5$ (lower panels)} \label{fig2}
\end{figure}

From Eq. (\ref{3.24}) using Eq. (\ref{4.14A}) we get
\begin{equation}
m(r)=\frac{r \left(16 \pi  \left(3+r^2 \Lambda  (3+\omega
)\right)+\left(3+15 \omega +4 r^2 (\Lambda +3 \Lambda  \omega
)\right) \varpi\right)}{12 \left(4 \pi  (3+\omega )+(1+3 \omega )
\varpi\right)}+d,
\end{equation}
where $d$ is the integrating constant.

\section{Physical features of compact stars in $f(\mathbb{T}, \mathcal{T})$ gravity}
In this section, we study some physical features of the compact star, in order to examine the physical validity and stability of the system
in the $f(\mathbb{T},\mathcal{T})$ gravity.

\subsection{Energy conditions}
In this subsection, we consider the following inequalities to check all the energy conditions whether these are satisfied or not and can be provided as follows:
\begin{eqnarray}\label{5.1}
 (i)NEC: \rho^{eff}+p_r^{eff}\geq 0, \rho^{eff}+p_t^{eff}\geq 0,\cr
(ii)WEC: \rho^{eff}+p_r^{eff}\geq 0, \rho^{eff} \geq 0,
\rho^{eff}+p_t^{eff}\geq 0,\cr (iii)SEC:\rho^{eff}+p_r^{eff}\geq
0, \rho^{eff}+p_r^{eff}+2p_t^{eff}\geq 0.\cr
\end{eqnarray}

This is important to note that for the physical validity of the stellar configuration an isotropic fluid sphere, specifically composed of SQM, should satisfy the above mentioned energy conditions at all the interior points of the system~\cite{Maurya2017}. In Fig. 3 we have plotted the energy conditions which shows the expected physical features.

\begin{figure}[!htpb]\centering
\includegraphics[width=6cm]{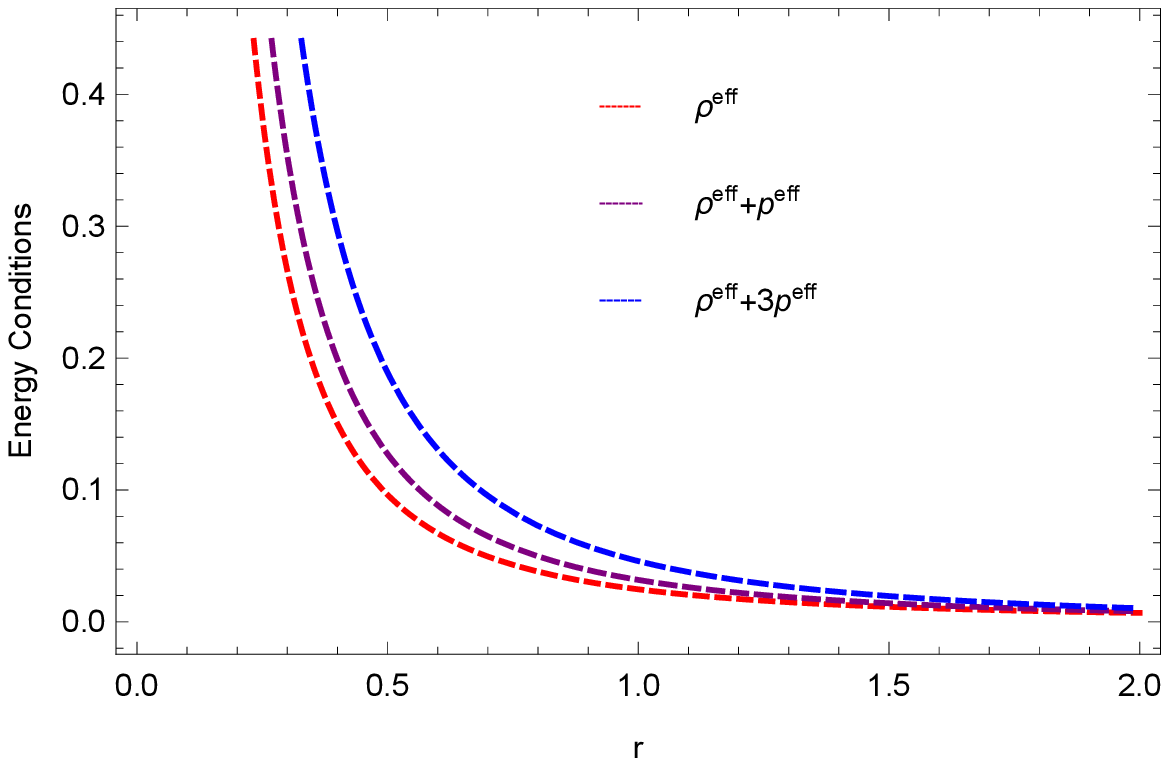}\\
\includegraphics[width=6cm]{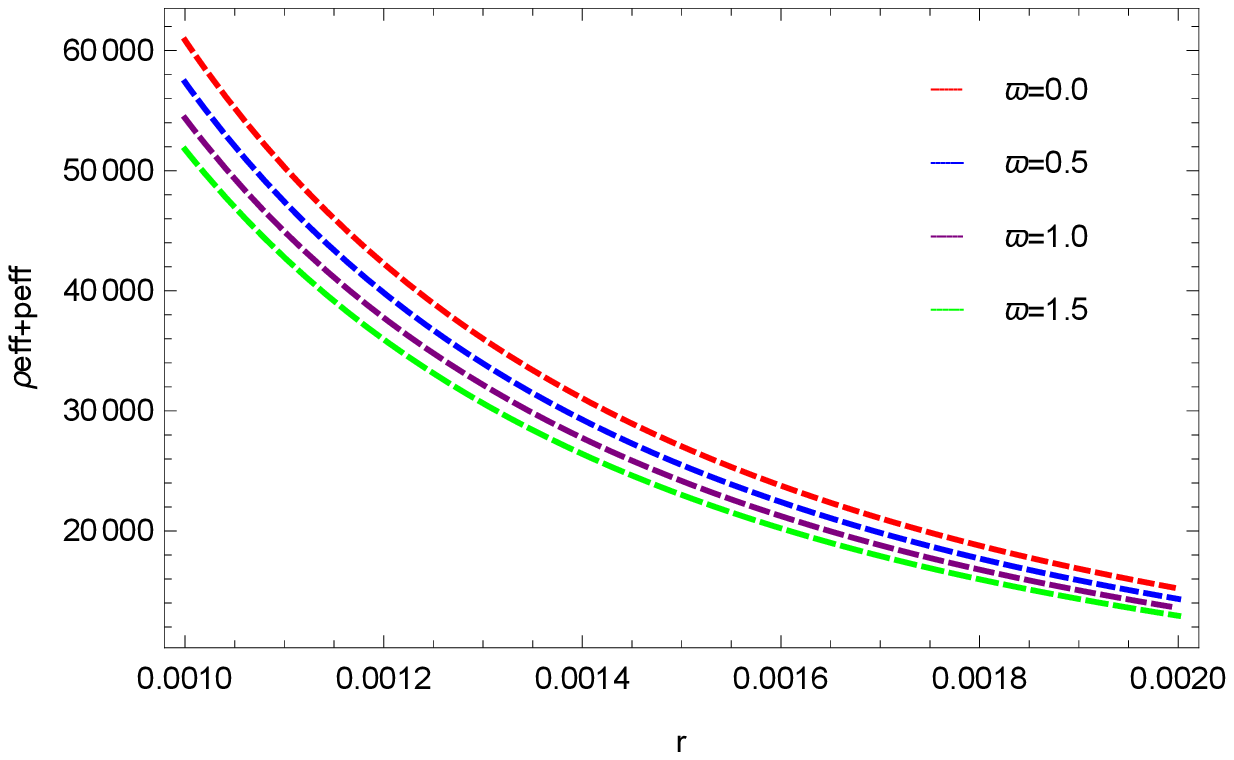}
\includegraphics[width=6cm]{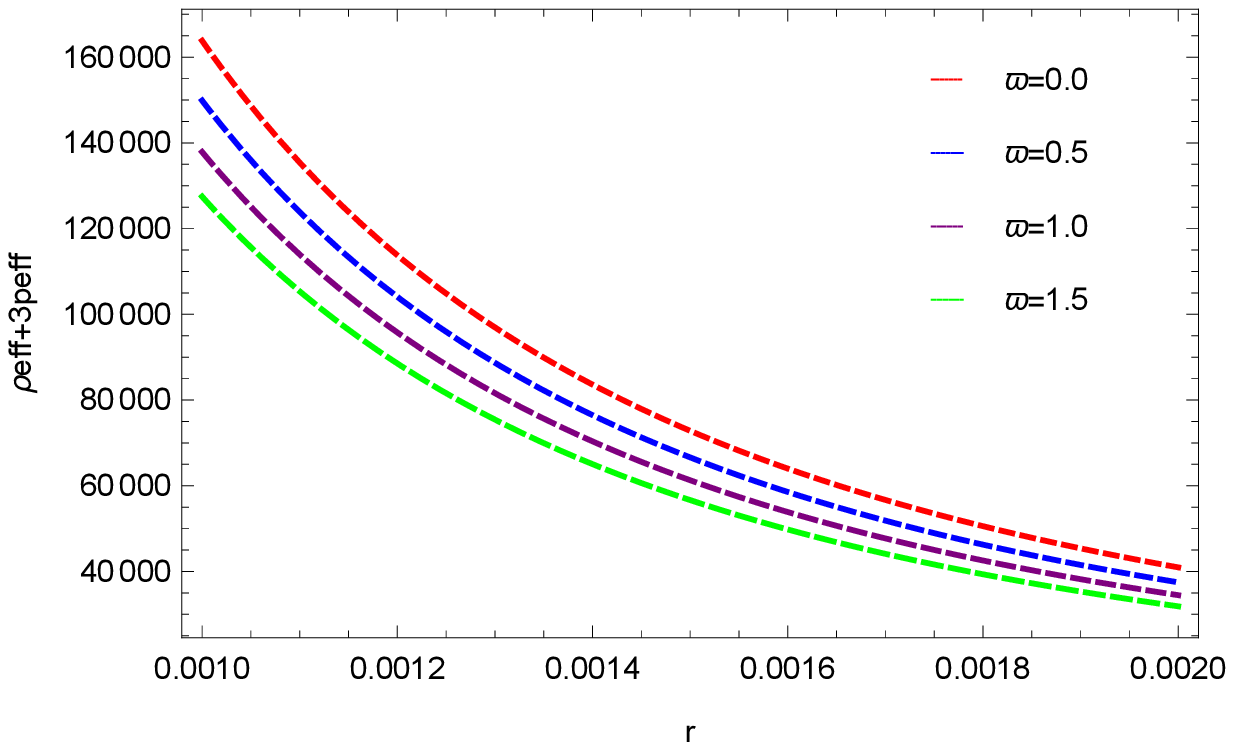}
\caption{Variations of all the energy conditions (upper panel) while variations of $\rho^{eff} + p^{eff}$ ($km^{-2}$) and $\rho^{eff}
+3p^{eff}$ ($km^{-2}$) are shown with respect to the radial coordinate $r$ (km) for $\varpi=0.0,0.5,1.0,1.5$ (lower panels).} \label{fig3}
\end{figure}

\subsection{Mass-radius relation}
The mass function within the radius r is given by
\begin{equation}\label{4.16A}
M(r)=\int_0^r 4\pi \acute{r}\rho^{eff}  d\acute{r}=\frac{1}{12} r
\left(4 r^2 \Lambda+\frac{3 \left(16 \pi +\varpi+5 \omega
\varpi\right)}{4 \pi (3+\omega )+(1+3 \omega ) \varpi}\right) 
\end{equation}

It is to note that Buchdahl~\cite{Buchdahl1959} prescribed restriction on the upper bound of the mass to radius ratio in uncharged perfect fluid model which is $\frac{2M}{R}<\frac{8}{9}$. Based on Eq. (\ref{4.16A}) Fig. 4 is the representative of the Mass-Radius relationship.  

\begin{figure}[!htpb]\centering
\includegraphics[width=6cm]{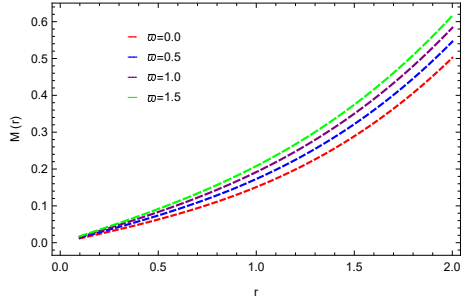}
\caption{Profile of the mass function $M(r)$ (km) with respect to the radial coordinate $r$ (km) for $\varpi=0.0,0.5,1.0,1.5$. } \label{fig4}
\end{figure}

\subsection{Stability Analysis}

\subsubsection{Modified TOV equation in $f(\mathbb{T},\mathcal{T})$ gravity theory}
It is to note that based on the Einstein field equations Oppenheimer and Volkoff~\cite{Oppenheimer1939} derived a differential equation extending the work of Tolman~\cite{Tolman1939} that describes the stellar structure of a compact object of static, isotropic material in hydrostatic equilibrium which is well known as the TOV equation. Rewriting the conservation Eq. (\ref{eq40}) for isotropic fluid distribution we have generalized the Tolman-Oppenheimer-Volkoff (TOV) equation as
\begin{equation}
-\frac{\nu'}{2} (p^{eff}+\rho^{eff})-\frac{dp^{eff}}{dr}+\frac{\varpi}{\Big(16\pi + 2\varpi\Big)}\Bigg\{(\rho^{eff})'-5
(p^{eff})' \Bigg\}=0.
\end{equation}

Here one can get the usual form of TOV equation in the case of general relativity if $\varpi=0$, however the above TOV equation in general can be described as
\begin{equation}
F_g+F_h+F_x=0,
\end{equation}
where $F_g$ is the gravitational force, $F_h$ is the hydrostatic force and $F_x$ is the additional force due to the modification of the gravitational Lagrangian of the standard Einstein-Hilbert action and they are defined as
\begin{eqnarray}
F_g=-\frac{\nu'}{2} (p^{eff}+\rho^{eff}),\cr
F_h=-\frac{dp^{eff}}{dr},\cr
F_x= \frac{\varpi}{\Big(16\pi + 2\varpi\Big)}\Bigg\{(\rho^{eff})'-5 (p^{eff})' \Bigg\}.
\end{eqnarray}

\begin{figure}[!htpb]\centering
\includegraphics[width=6cm]{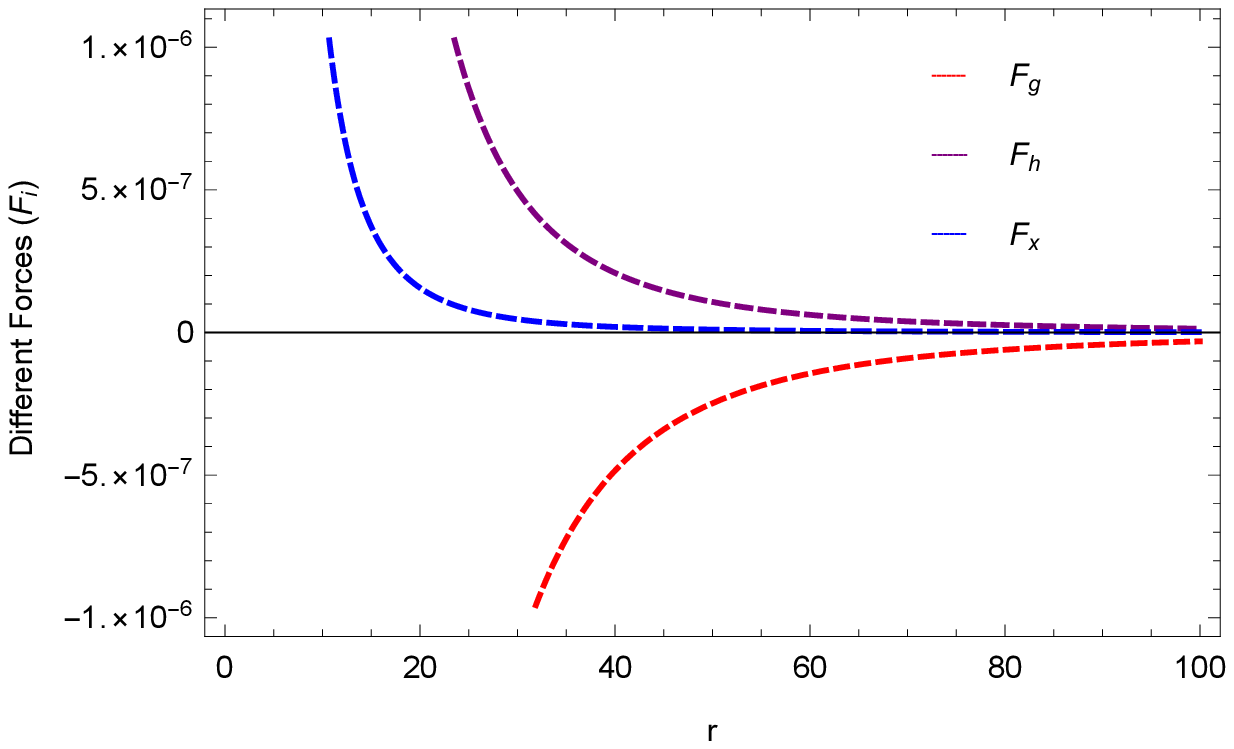}
\includegraphics[width=6cm]{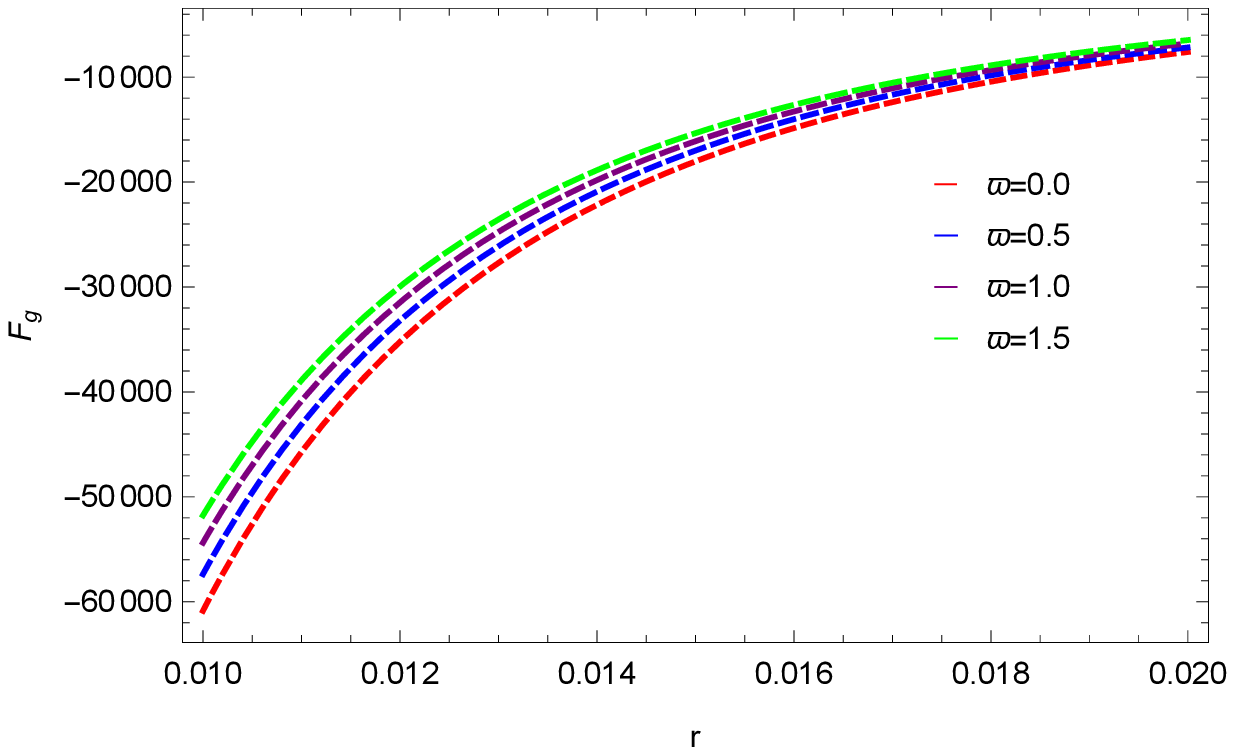}
\includegraphics[width=6cm]{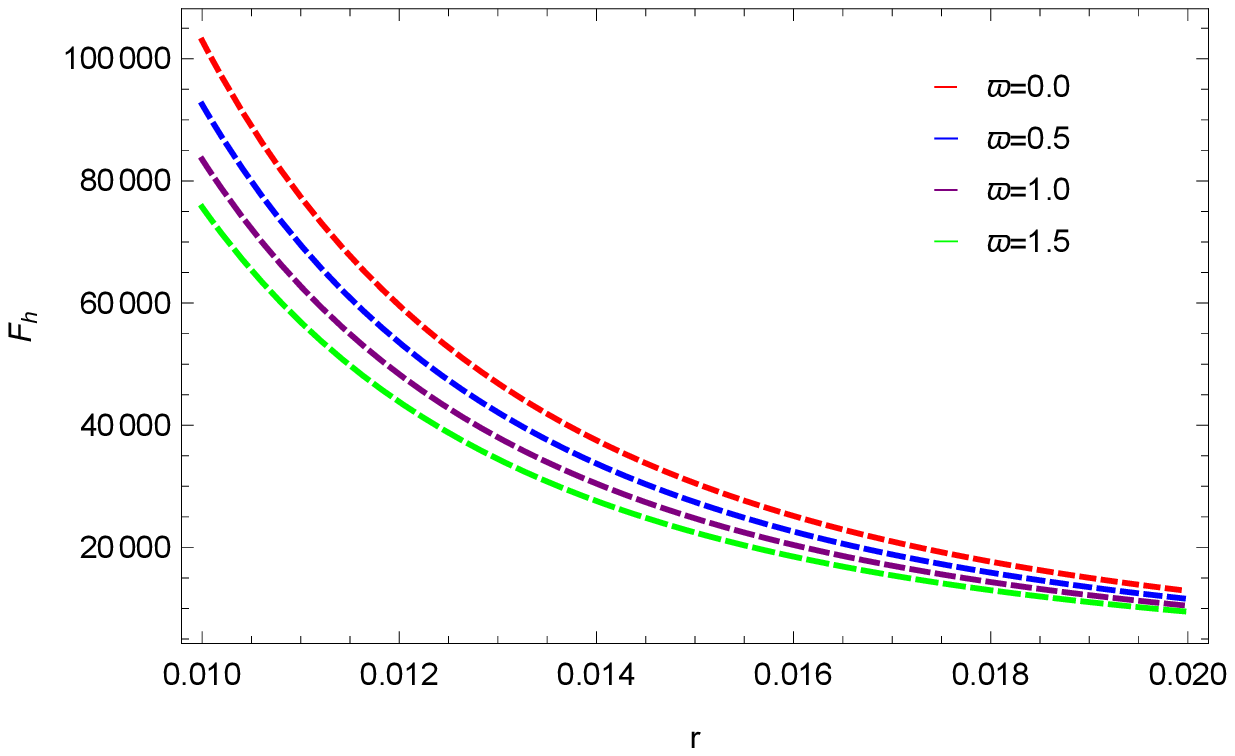}
\includegraphics[width=6cm]{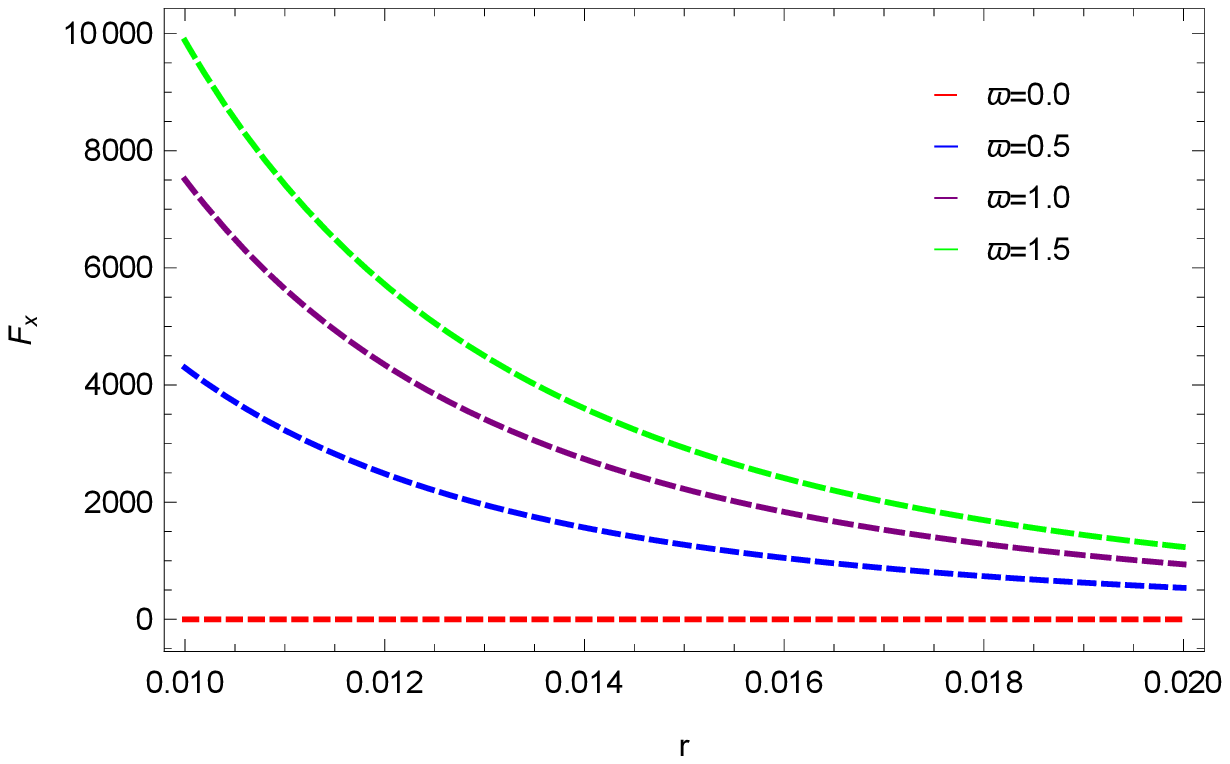}
\caption{Profiles of the three different forces, viz. the gravitational force $(F_g)$, the hydrostatic force $(F_h)$ and the additional
force $(F_x)$ are plotted against $r$ (km) in general (upper left panel) while their variations for $\varpi=0.0,0.5,1.0,1.5$ respectively (upper right and lower panels).} \label{fig5}
\end{figure}

We have shown in Fig. 5 the balancing features of different forces to make the configuration as stable one.

\subsubsection{Speed of Sound}
To apply the Herrera's cracking concept and the causality condition~\cite{Herrera1992} we now define the square of the sound speed for the present isotropic matter distribution which can be given as
\begin{eqnarray}
v_s^{2}=\frac{dp^{eff}}{d\rho^{eff}}=\frac{\varpi+16 \pi  \omega -3 \varpi \omega }{16 \pi +\varpi+5 \varpi \omega }.
\end{eqnarray}

In our study, the variation of $v^2_{s}$ w.r.t. radial coordinate $r$ has been featured in Fig.~6 which clearly shows that value of $v^2_{s}$ remains within the range (0-1) in concordance to the causality condition.

\begin{figure}[!htpb]\centering
\includegraphics[width=6cm]{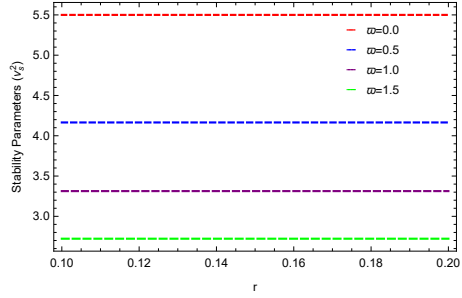}
\caption{Variation of square of the sound speed is plotted with respect to the radial coordinate $r$ (km) for $\varpi=0.0,0.5,1.0,1.5$. } \label{fig6}
\end{figure}

\subsection{Compactification factor and surface redshift}
In the present case of $f(\mathbb{T},\mathcal{T})$ gravity theory the compactification factor is given by
\begin{eqnarray}
u(r)=\frac{M(r)}{r} =\frac{1}{12}  \left(4 r^2 \Lambda+\frac{3
\left(16 \pi +\varpi+5 \omega \varpi\right)}{4 \pi (3+\omega
)+(1+3 \omega ) \varpi}\right).
\end{eqnarray}

Therefore, in terms of the compactification factor we can now define the surface redshift as
\begin{eqnarray}
Z_s=(1-2u)^{-\frac{1}{2}}-1= -1+\frac{\sqrt{6}}{\sqrt{-4 r^2 \Lambda +\frac{3 (8 \pi +\varpi) (1+\omega )}{\varpi+3 \varpi
\omega +4 \pi (3+\omega )}}}.
\end{eqnarray}

\begin{figure}[!htpb]\centering
\includegraphics[width=6cm]{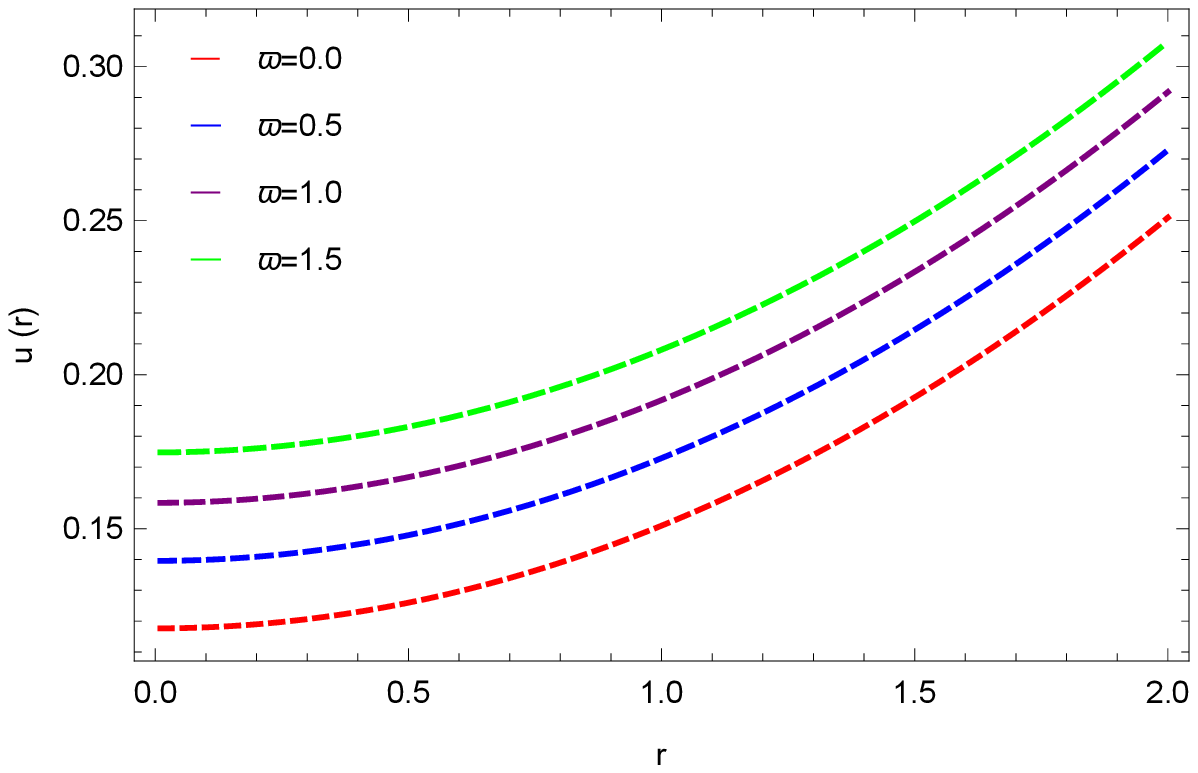}
\includegraphics[width=6cm]{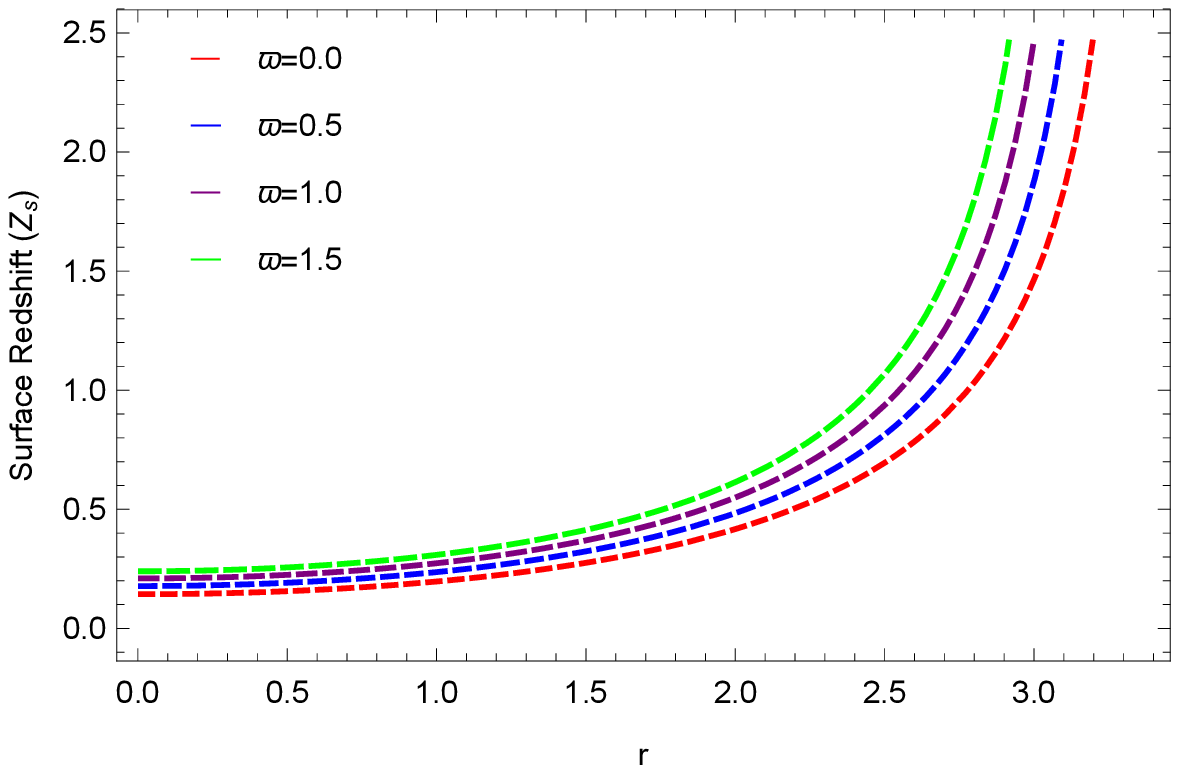}
\caption{The compactness factor $u(r)$ and surface redshift $(Z_s)$ are plotted with respect to the radial coordinate $r$ (km) for $\varpi=0.0,0.5,1.0,1.5$. } \label{fig7}
\end{figure}

Figs. 7 shows the compactification factor and surface redshift, respectively, in the left and right panels.

\section{Discussion and conclusion}
In the present work our motivation was to investigate compact stars, specifically quark stars, under $f(\mathbb{T},\mathcal{T})$ gravity where we have basically considered a different form in r.h.s. of the field equations in terms of torson scalar and trace of the energy-momentum tensor. To keep the model simplest one, we also consider the spacetime as spherical and static. Accordingly, unlike the previous investigation~\cite{Salako2020} here we have exempted anisotropy and also have exploited the technique of CKV to  make the nonlinear equations easily solvable. The solutions obtained are interesting as far as different physical checklists via graphical plots are concerned and can be put in a brief as follows:

(1) In Fig. 1 we have plotted metric potentials with respect to $r$ for $\varpi=0.0,0.5,1.0,1.5$ which exhibit usual physical features, i.e. both the metric potentials have finite values at the centre and they monotonically increase from the centre to the surface. 

(2) In Fig. 2, the effective density and pressure profiles are also interesting which drop down from very high values to the low values. In other words, this figure features that both $\rho^{\textit{eff}}$ and $p^{\textit{eff}}$ have maximum values at the centre and they gradually decrease to reach the minimum value at the surface and validates the physical viability of the obtained solutions. This also confirms that our system is free from any singularities, viz., either geometrical or physical singularity. 

(3) We have plotted the energy conditions in Fig. 3 and note that all the conditions, viz. NEC, WEC and SEC have fulfilled the physical criteria for our obtained solution.

(4) Based on Eq. (\ref{4.16A}) we have plotted the Mass-Radius relationship in Fig. 4 and observe that there is a central core of the star with a finite radius of~8 km. It is also notable that $M(r)/r \le 0.44$ as the condition imposed by Buchdahl~\cite{Buchdahl1959}.

(5) We have analyzed stability of the model star under the heads (i) TOV equations and (ii) sound speed. From Fig. 5 one can note that the forces under action balance each other to make the spherical distribution stable. In this context its worthy to mention that according to Jasim et al.~\cite{Jasim2020} the stellar properties of any compact object are dependent on its internal structure, which is described by the EOS. The stability of the compact stellar system can be obtained when the inward gravitational force should be counter balanced by the repulsive and outward forces produced inside the stellar object in such a way that the resultant force on the system would be zero. They argue that this is essential as otherwise a small perturbation will cause the system to be unstable. On the other hand, Fig. 6 indicates that the prescribed condition of sound speed, i.e. $0<v^2_s<1$, is maintained.

(6) In Figs. 7 we have shown the behaviour of the compactification factor and surface redshift (left and right panels) both are exhibiting physically viable features. In this connection we are interested to mention that the compactification factor for a static, spherically symmetric, perfect fluid star classifies the stellar objects into different categories as follows~\cite{Jotania2006}: (i) normal star: $M/R\sim 10^{-5}$, (ii) white dwarf: $M/R\sim 10^{-3}$, (iii) neutron star: $10^{-1}<M/R<1/4$, (iv) ultra-dense compact star: $1/4<M/R<1/2$ and (v) black hole: $M/R=1/2$. On the other hand, according to Barraco and Hamity~\cite{Barraco2002} for an isotropic star the redshift must be $\leq 2$, provided the Cosmological constant is absent. Therefore, from Fig. 7 it is evident that the model represents a stable and ultra-dense strange star.

(7) In all the Figs. 1 - 7, one can note that $\varpi=0.0$ represents the GR case while the other values, viz., $\varpi=0.5,1.0,1.5$ are representatives of the $f(\mathbb{T},\mathcal{T})$ gravity theory. The results of the different plots of the physical parameters demonstrate that non-zero values of $\varpi$ have definite impact on the distribution of the fluid sphere, e.g. in Fig. 7, as we go on increasing the value of $\varpi$ both the physical parameters, i.e. the compactification factor as well as surface redshift get increased in their respective values. Therefore, specifically the higher values of $\varpi$ provide more condense strange star. Thus, overall inspection indicates that there is a direct proportionality between the modified gravity parameter and physical parameters of the model. This feature is therefore a distinct difference between Einstein's GR and its modified theory and hence provides justification for considering modification in GR in the present model. 

To conclude we have found a new type of stable compact configuration, which can be used for confrontation with the observational data,
extending astrophysical probes for models of modified gravity.

\section*{Acknowledgement} CR and SR thank IUCAA, Pune, India for hospitality and support during an academic visit where a part of this work is accomplished. The research by M.K. was financially supported by Southern Federal University, 2020 Project VnGr/2020-03-IF.


\begin{thebibliography}{99}

\bibitem{Schwarzschild1916a} K. Schwarzschild, Sitz. Preu. Akad. Wissenchaften, Berlin Phys. Math. Klasse 189 (1916).
    
\bibitem{Schwarzschild1916b} K. Schwarzschild, Sitz. Preu. Akad. Wissenchaften, Berlin Phys. Math. Klasse 424 (1916).

\bibitem{Oppenheimer1939} J.R. Oppenheimer, G.M. Volkoff, Phys. Rev. D {\bf 55} (1939) 374.

\bibitem{Tolman1939} R.C. Tolman, Phys. Rev. D {\bf 55} 364 (1939).

\bibitem{Delgaty1998} M.S.R. Delgaty, K. Lake, Comput. Phys. Commun. {\bf 115} (1998) 395.

\bibitem{Bombaci2004} I. Bombaci, I. Parenti, I. Vidana, Astrophys. J. {\bf 614} (2004) 314.

\bibitem{Staff2007} J. Staff, R. Ouyed, M.A. Bagchi, Astrophys. J. {\bf 667} (2007) 340.

\bibitem{Herzog2011} M. Herzog, F.K. Ropke, Phys. Rev. D {\bf 84} (2011) 083002.

\bibitem{Wheeler1962} J.A. Wheeler, Geometrodynamics (Academic Press, New York, p. 25, 1962).

\bibitem{Kamenschik2001} A. Kamenschik, U. Moschella, V. Pasquier, Phys. Rev. Lett. B {\bf 511} (2001) 265.

\bibitem{Padmanabhan2002} T. Padmanabhan, T.R. Chaudhury, Phys. Rev. D {\bf 66} (2002) 081301.

\bibitem{Bento2002} M.C. Bento, O. Bertolami, A.A. Sen, Phys. Rev. D {\bf 66} (2002) 043507.

\bibitem{Caldwell2002} R.R. Caldwell, Phys. Lett. B {\bf 545} (2002) 23.

\bibitem{Nojiri2003a} S. Nojiri, S.D. Odintsov, Phys. Lett. B {\bf 562} (2003) 147.

\bibitem{Nojiri2003b} S. Nojiri, S.D. Odintsov, Phys. Lett. B {\bf 565} (2003) 1.

\bibitem{Riess2004} A.G. Riess, et al. Astrophys. J. {\bf 607} (2004) 665.

\bibitem{Eisenstein2005} D.G. Eisenstein, et al.: Astrophys. J. {\bf 633} (2005) 560.

\bibitem{Astier2006} P. Astier, et al.: Astron. Astrophys. {\bf 447} (2006) 31.

\bibitem{Spergel2007} D.L. Spergel, et al.: Astrophys. J. Suppl. {\bf 170} (2007) 377.

\bibitem{Nojiri2003} S. Nojiri, S.D. Odintsov, Phys. Rev. D {\bf 68} (2003) 123512.

\bibitem{Carroll2004} S.M. Carroll, V. Duvvuri, M. Trodden, M.S. Turner, Phys. Rev. D {\bf 70} (2004) 043528.

\bibitem{Allemandi2003} G. Allemandi, A. Borowiec, M. Francaviglia, S.D. Odintsov, Phys. Rev. D {\bf 72} (2003) 063505.

\bibitem{Nojiri2007} S. Nojiri, S.D. Odintsov, Int. J. Geom. Method. Mod. Phys. {\bf 04} (2007) 115.

\bibitem{Bertolami2007} O. Bertolami, C.G. Bohmer, T. Harko, F.S.N. Lobo, Phys. Rev. D {\bf 75} (2007) 104016.

\bibitem{Bengachea2009} G.R. Bengochea, R. Ferraro, Phys. Rev. D {\bf 79} (2009) 124019.

\bibitem{Linder2010} E.V. Linder, Phys. Rev. D {\bf 81} (2010) 127301.

\bibitem{Bamba2010a} K. Bamba, C.Q. Geng, S. Nojiri, S.D. Odintsov, Eur. Phys. Lett. {\bf 89} (2010) 50003.

\bibitem{Bamba2010b} K. Bamba, S.D. Odintsov, L. Sebastiani, S. Zerbini, Eur. Phys. J. C {\bf 67} (2010) 295.

\bibitem{Rodrigues2010} M.E. Rodrigues, M.J.S. Houndjo, D. Momeni, R. Myrzakulov, Can. J. Phys. {\bf 92} (2014) 173.

\bibitem{Harko2011}T. Harko, F.S.N. Lobo, S. Nojiri, S.D. Odintsov, Phys. Rev. D {\bf 84} (2011) 024020.

\bibitem{Harko2008} T. Harko, Phys. Lett. B {\bf 669} (2008) 376.

\bibitem{Bisabr2012} Y. Bisabr, Phys. Rev. D {\bf 86} (2012) 044025.

\bibitem{Jamil2012} M. Jamil, D. Momeni, R. Muhammad, M. Ratbay, Eur. Phys. J. C {\bf 72} (2012) 1999.

\bibitem{Alvarenga2013} F.G. Alvarenga, A. de la Cruz-Dombriz, M.J.S. Houndjo, M.E. Rodrigues, D. Saez-Gomez, Phys. Rev. D {\bf 87} (2013) 103526.

\bibitem{Shabani2013} H. Shabani, M. Farhoudi, Phys. Rev. D {\bf 88} (2013) 044048.

\bibitem{Shabani2014} H. Shabani, M. Farhoudi, Phys. Rev. D {\bf 90} (2014) 044031.

\bibitem{Zaregonbadi2016} R. Zaregonbadi, M. Farhoudi, Gen. Rel. Grav. {\bf 48} (2016) 142.

\bibitem{Shabani2017a} H. Shabani, Int. J. Mod. Phys. D {\bf 26} (2017) 1750120.

\bibitem{Shabani2017b} H. Shabani, A.H. Ziaie, Eur. Phys. J C {\bf 77} (2017) 31.

\bibitem{Junior2014} E.L.B. Junior, M.E. Rodrigues, I.G. Salako, M.J.S. Houndjo, Class. Quant. Gravit. {\bf 33} (2014) 125006.

\bibitem{Salako2015} I.G. Salako, A. Jawad, S. Chattopadhyay, Astrophys. Sp. Sci. {\bf 358} (2015) 13.

\bibitem{Ganiou2016a} M. G. Ganiou, Ines G. Salako, M. J. S. Houndjo and J. Tossa, Int. J. Theor. Phys. {\bf 55} (2016) 3954.

\bibitem{Ganiou2016b} M. G. Ganiou, Ines G. Salako, M. J. S. Houndjo and  J. Tossa, Astrophys. Space Sci. {\bf 361} (2016) 57.

\bibitem{Salako2017} Ines G. Salako,  A. Jawad and H. Moradpour, Int. J. Geom. Meth. Mod. Phys. {\bf 15} (2017) 1850063.

\bibitem{Ghosh2020} S. Ghosh, A. D. Kanfon, A. Das,  M. J. S. Houndjo, Ines G. Salako and S. Ray, Int. J. Mod. Phys. A {\bf 35} (2020) 2050017.

\bibitem{Nojiri2005} S. Nojiri, S.D. Odintsov, Phys. Lett. B {\bf 631} (2005) 1.

\bibitem{Ruderman1972} R. Ruderman, Ann. Rev. Astron. Astrophys. {\bf 10} (1972) 427.

\bibitem{Herrera1997} L. Herera, N.O. Santos, Phys. Rep. {\bf 286} (1997) 53.

\bibitem{Hossaein2012} S.M. Hossaein, F. Rahaman, J. Naskar, M. Kalam, S. Ray, Int. J. Mod. Phys. D {\bf 21} (2012) 1250088.

\bibitem{Kalam2014} M. Kalam, F. Rahaman, S. Molla, S.M. Hossein, Astrophys. Sp. Sci. {\bf 349} (2014) 865. 

\bibitem{Bhar2015} P. Bhar, Astrophys. Sp. Sci. {\bf 356} (2015) 309.

\bibitem{Abbas2015} G. Abbas, M. Zubair, G. Mustafa, Astrophys. Sp. Sci. {\bf 358} (2015) 26.

\bibitem{Rahaman2014} F. Rahaman, K. Chakraborty, P.K.F. Kuhfittig, G.C. Shit, M.A. Rahman, Eur. Phys J. C {\bf 74} (2014) 3126.

\bibitem{Abranil2016} J.D.V. Abranil, M. Malheiro, J. Cosmol. Astropart. Phys. {\bf 11} (2016) 012.

\bibitem{Murad2016} M.H. Murad, Astrophys. Sp. Sci. {\bf 361} (2016) 20.

\bibitem{Oliveira2015} A.M. Oliveira, H.F.S. Velten, J.C. Fabris, L. Casarini, Phys. Rev. D {\bf 92} (2015) 044020.

\bibitem{Sarif2018} M. Sarif, A. Waseem, Eur. Phys. J. C {\bf 78} (2018) 868.

\bibitem{Saha2018} P. Saha, U. Debnath, Adv. High Energy Phys. {\bf 3901790} (2018).

\bibitem{Maurya2019} S.K. Maurya, F. Tello-Ortiz, Eur. Phys. J. C {\bf 79} (2019) 85.

\bibitem{Prasad2019} A.K. Prasad, J. Kumar, S.K. Maurya, B. Dayanandan, Astrophys. Sp. Sci. {\bf 364} (2019) 66.

\bibitem{Saha2019} P. Saha, U. Debnath, Eur. Phys. J. C {\bf 79} (2019) 919.

\bibitem{Abbas2019} G. Abbas, M.R. Shahzad, Astrophys. Sp. Sc. {\bf 364} (2019) 50.

\bibitem{Shahzad2019} M.R. Shahzad, G. Abbas, Int. J. Geom. Meth. Phys. {\bf 16} (2019) 1950132.

\bibitem{Naxar2020} H. Nazar, G. Abbas, Chin. J. Phys. {\bf 63} (2020) 436.

\bibitem{Rahaman2012} F. Rahaman, R. Sharma, S. Ray, R. Maulick, I. Karar, Eur. Phys. J. C {\bf 72} (2012) 2071.

\bibitem{Deb2018a} D. Deb, B.K. Guha, F. Rahaman, S. Ray, Phys. Rev. D {\bf 97} (2018) 084026.

\bibitem{Deb2018b} D. Deb, F. Rahaman, S. Ray, B.K. Guha, J. Cosmol. Astropart. Phys. {\bf 03} (2018) 044.

\bibitem{Salako2020} I.G. Salako, M. Khlopov, S. Ray, M.Z. Arouko, P. Saha, U. Debnath, Universe {\bf 6} (2020) 167.

\bibitem{Deb2016} D. Deb, S. Roy Chowdhury, B.K. Guha, S. Ray, arxiv 2016, arxiv 1611.2053.

\bibitem{Chodos1974} A. Chodos, R.L. Jaffe, K. Johnson, C.B. Thorn, V.F. Weisskopf, Phys. Rev. D {\bf 9} (1974) 3474.

\bibitem{Harko2014} T. Harko, F.S.N. Lobo, G. Otalora, E.N. Saridakis, J. Cosmol. Astropart. Phys. {\bf 12} (2014) 021.

\bibitem{Aziz2019} A. Aziz, S. Ray, F. Rahaman, M. Khlopov, B.K. Guha, Int. J. Mod. Phys. D {\bf 28} (2019) 1941006. 

\bibitem{Jasim2020} M.K. Jasim, S.K. Maurya, S. Ray, D. Shee, D. Deb, F. Rahaman, Resul. Phys. {\bf20} (2021) 103648. 
 
\bibitem{Biswas2021} S. Biswas, D. Deb, S. Ray, B.K. Guha, Anisotropic charged strange stars in Krori-Barua spacetime under $f(R,\mathcal{T})$ gravity, {\it Accepted in Ann. Phys.} (2021).

\bibitem{Daouda2011} M.H. Daouda, M.E. Rodrigues and M.J.S. Houndjo, Eur.Phys.J. C {\bf 71} (2011) 1817.

\bibitem{Daouda2012} M.H. Daouda, M.E. Rodrigues, M.J.S. Houndjo, Eur. Phys. J. C {\bf 72} (2012) 1890.   

\bibitem{Zdunik2000} J.L. Zdunik, T. Bulik, W. Kluzniak, P. Haensel, D. Gondek-Rosinska, Astron. Astrophys. \textbf{359} (2000) 143.

\bibitem{Maieron2004} C. Maieron, M. Baldo, G.F. Burgio, H.J. Schulze, Phys. Rev. D \textbf{70} (2004) 043010.

\bibitem{Nicotra2006} O.E. Nicotra, M. Baldo, G.F. Burgio, H.-J. Schulze, Phys. Rev. D \textbf{74} (2006) 123001.

\bibitem{Bao2009} T. Bao, G.-Z. Liu, M.-F. Zhu, Chin. Phys. C \textbf{33} (2009) 340.

\bibitem{Uechi2010} S.T. Uechi, H. Uechi, arXiv:1003.4815 [nucl-th].

\bibitem{Isayev2015} A.A. Isayev, Phys. Rev. C \textbf{91} (2015) 015208.

\bibitem{Cardoso2017} P.H.G. Cardoso, T.N. da Silva, A. Deppman, D.P. Menezes, Eur. Phys. J. A \textbf{53} (2017) 191.

\bibitem{Joshi2020} S. Joshi, S. Sau, S. Sanyal, arXiv:2002.07647 [nucl-th].

\bibitem{Harko2002} M.K. Mak, T. Harko, Proc. Roy. Soc. Lond. A \textbf{459} (2003) 393.  

\bibitem{Maurya2017} S.K. Maurya, Y.K. Gupta, S. Ray, D. Deb, Eur. Phys. J. C {\bf77} (2017) 45.

\bibitem{Buchdahl1959} H.A. Buchdahl, Phys. Rev. D {\bf116} (1959) 1027.

\bibitem{Herrera1992} L. Herrera, Phys. Lett. A {\bf165} (1992) 206.

\bibitem{Jotania2006} K. Jotania and R. Tikekar, Int. J. Mod. Phys. D {\bf15}, 1175 (2006)

\bibitem{Barraco2002} D.E. Barraco and V.H. Hamity, Phys. Rev. D {\bf65}, 124028 (2002)


\end{thebibliography}
\end{document}